# Some Preliminary Steps Towards Metaverse Logic

Antonio L. Furtado, Marco A. Casanova, Edirlei Soares de Lima

PUC-Rio, Departamento de Informática, Rio de Janeiro, Brasil
{furtado,casanova}@inf.puc-rio.br
Universidade Europeia, Faculty of Design, Technology and Communication – IADE, Lisbon, Portugal
edirlei.lima@universidadeeuropeia.pt

**Abstract:** Assuming that the term 'metaverse' could be understood as a computer-based implementation of multiverse applications, we started to look in the present work for a logic that would be powerful enough to handle the situations arising both in the real and in the fictional underlying application domains. Realizing that first-order logic fails to account for the unstable behavior of even the most simpleminded information system domains, we resorted to non-conventional extensions, in an attempt to sketch a minimal composite logic strategy. The discussion was kept at a rather informal level, always trying to convey the intuition behind the theoretical notions in natural language terms, and appealing to an AI agent, namely ChatGPT, in the hope that algorithmic and common-sense approaches can be usefully combined.

## 1. Introduction

The current notion of *metaverse*, referring to the exploitation of virtual worlds[1], has a broad scope, as indicated in a recent report [Zhu], presented on August 26, 2022, by the Congressional Research Service[2]:

> The term metaverse has been in use since at least 1992 and currently it generally refers to the concept of an immersive and persistent virtual world where users can communicate and interact with other users and the surrounding environment and engage in social activities, similar to interactions in the physical world … This report uses the term "the metaverse" to refer to the general concept and its related technologies but not a specific configuration of devices, platforms, applications, and services[3].

The report came as a response to the United States Congress growing interest on certain computer-based technologies becoming available to internet users. In its introductory paragraph, some of these technologies are listed, and – more importantly to the objective of the present study – the term metaverse is associated with *immersive* virtual worlds:

> Congress has long paid attention to technologies for users to access computer-simulated environments and participate in virtual activities on the internet. These technologies include augmented reality, mixed reality, and virtual reality technologies (AR, MR, and VR respectively) that show potential for innovation in a variety of applications such as entertainment, healthcare, engineering, real estate, retail, military, education, and collaborative work. AR, MR, and VR technologies, some enthusiasts argue, may support new ways for users to interact, work, socialize, transact, and access services in an *immersive virtual world*, which is often referred to as the metaverse.

We shall assume here that virtual worlds would more effectively offer to users an immersive environment if they met the plausibility requirements of *possible worlds*, which would lead to treating metaverse with the help of rigorous formal approaches characterizing the notion of *multiverse*. Prominent among these is David Lewis's theory [Lewis 1], which relies on *counterfactual* logic reasoning [Lewis 2]. For him, worlds conceived to answer 'what if' questions are as real as the world where we currently dwell [Lewis 2]:

---

[1] An early science fiction inspiration for machine-induced 'virtual worlds': *Star Trek*'s 'holodeck' - https://en.wikipedia.org/wiki/Holodeck
[2] https://crsreports.congress.gov
[3] One well-known specific case being the "Facebook Metaverse", an online reality virtual space project at Meta Platforms (Meta Inc., formerly, Facebook Inc.).

> I believe there are possible worlds other than the one we happen to inhabit. If an argument is wanted, it is this: It is uncontroversially true that things might have been otherwise than they are ... I therefore believe in the existence of entities which might be called 'ways things could have been'. I prefer to call them 'possible worlds'.

While this reality claim has been questioned by some (e.g. [Stalnaker]), others, such as Max Erik Tegmark[4], a physicist, cosmologist and machine learning researcher and professor at MIT, go much beyond, arguing for the existence of a multiverse comprising four levels of parallel universes [Tegmark]:

> The frontiers of physics have gradually expanded to incorporate ever more abstract (and once metaphysical) concepts such as a round Earth, invisible electromagnetic fields, time slowdown at high speeds, quantum superpositions, curved space, and black holes. Over the past several years the concept of a multiverse has joined this list. It is grounded in well-tested theories such as relativity and quantum mechanics, and it fulfills both of the basic criteria of an empirical science: it makes predictions, and it can be falsified. Scientists have discussed as many as four distinct types [levels] of parallel universes.

The fourth level, involving 'other mathematical structures', is most intriguing:

> The initial conditions and physical constants in the Level I, Level II and Level III multiverses can vary, but the fundamental laws that govern nature remain the same. Why stop there? Why not allow the laws themselves to vary?

Indeed, even the walls between reality and magic appear to be crumbling down with the advancements of quantum mechanics. In 1890, James Frazer had thus described a primitive magic belief [Frazer]:

> The other great branch of sympathetic magic, which I have called Contagious Magic, proceeds upon the notion that things which have once been *conjoined* must remain ever afterwards, even when quite dissevered from each other, in such a sympathetic relation that *whatever is done to the one must similarly affect the other*.

Strangely enough, three physicists – Alain Aspect, John F. Clauser and Anton Zeilinger – won the 2022 Nobel Prize in Physics for their work on a physical phenomenon that can hardly be distinguished from this magic arcane[5]:

> Quantum entanglement is a bizarre, counterintuitive phenomenon that explains how two subatomic particles can be intimately *linked to each other* even if separated by billions of light-years of space. Despite their vast separation, *a change induced in one will affect the other*.

Since metaverse applications include entertainment, fictional narrative worlds should also be contemplated,[6] as in Marie-Laure Ryan's comparative analysis [Ryan]:

> In physics, the existence of parallel universes has been postulated on the cosmic level to describe what lies on the other side of black holes and, on the level of subatomic particles, to avoid the paradoxes of quantum mechanics. In *narratology*, the philosophical idea of a plurality of possible worlds and the contrast between the actual and the possible provide a model of the cognitive pattern into which readers organize information in order to interpret it as a story. But the many-worlds interpretation of physics and the possible worlds (PW) model of narrative differ in their conception of the ontological status of the multiple worlds: in physics they are all actual, while narrative theory stresses the contrast between *actuality* and *mere possibility*. This does not mean that the PW model is incompatible with the many-worlds cosmology proposed by physics: faced with a narrative that presents multiple realities as existing objectively, the theory would simply claim that the actual domain is made up of a number of different worlds and that the distinction actual/nonactual repeats itself within each of these parts.

One might imagine that only those merely possible worlds would demand special methods, such as counterfactual reasoning [Lewis 2], but even actual, practical worlds of business information systems [Gottin et al. 1; Gottin et al. 2], that might be involved in metaverse serious applications, suffer from a lack of stability that goes far beyond the reach of standard logic [Georgeff et al.]:

---

[4] https://en.wikipedia.org/wiki/Max_Tegmark
[5] https://www.space.com/31933-quantum-entanglement-action-at-a-distance.html
[6] In addition, note that the term 'metaverse' occurred for the first time in a science fiction novel, *Snow Crash*, written by Neal Stephenson, and published in 1992.

Most applications of computer systems are algorithmic, working with perfect information. But in a highly competitive world, businesses need systems that are much more complex than this — systems that are embedded in a *changing world*, with access to only partial information, and where uncertainty prevails. Moreover, the frequency with which the behaviour of these systems needs to be changed (as new information comes to light, or new competitive pressures emerge), is increasing dramatically. As we all know, but seem not to have fully understood (at least in the way physicists have) the world is complex and dynamic, a place where chaos is the norm, not the exception.

Accordingly, we shall try in the present study to sketch what might be developed, after much extensive and careful research work, into a composite logic strategy – comprising, besides counterfactuals, other non-standard extensions – to handle the plurality of worlds underlying metaverse projects.

Section 2 deals with the convenience to extend first-order logic to cope with world-state evolution, and outlines the structure of the proposed strategy. Section 3 is rather atypical: it was not written by us, and should instead be credited to an AI agent, namely ChatGPT, to which we asked whether or not it was familiar with each recommended logic extension and, in case of a positive answer, to briefly describe it and supply an explanatory example. Section 4 describes our initial attempt to start a two-way communication between ChatGPT and Prolog, with the code developed to run the examples being included in the appendixes. Section 5 contains concluding remarks.

**2. Towards a composite logic strategy**

As a tentative consequence of what was said in the previous section, we shall understand metaverse as a computer implementation of a given multiverse, supporting serious or entertaining applications over that multiverse. And now, for simplicity, we shall consider that each universe in the underlying multiverse involves a single application domain, for which a conceptual specification has been provided. Such specification should include, as a critical part, a set of rules, called *integrity constraints* in database parlance, whose purpose would be to determine the existence and behavior of the domain's constituent instances.

A common practice, widely adopted since the proposal of the relational model, has been to express such rules as first-order logic sentences. Despite the recognized power of first-order logic, the unstable nature of real world domains, pointed out in the previous section [Georgeff et al.], makes it unsuitable to handle seemingly atypical situations.

Roughly speaking, traditional logic relies on certain assumptions, which are often unrealistic:

- **Non contradiction** – sentences cannot be both true and not true.
- **Monotonicity** – present conclusions are not invalidated by new information.
- **Closure** – the domain's model is complete – whatever is outside it is either false or irrelevant. While making provision for non-monotonic state updates, this leads to the notion of 'closed world assumption' (in databases), and to 'negation as failure' (in logic programming).
- **Deductive reasoning** – given a rule $A \rightarrow B$, then, if fact A is true, fact B is necessarily true.
- **Determinism** – given an event-producing operation O with preconditions C and effects E, if C holds and O takes place, then E will necessarily hold, which disallows alternative specification of the E effects, as well as the possibility that the execution of O may partially or totally fail.
- **Awareness** – agents know all facts, rules and operations of information systems they participate in, and are competent to evaluate their current situation.
- **Rationality** – agents are motivated by goals that improve their situation, and to reach a goal they utilize plans that are executable sequences of event-producing operations.

In the world of politics, the first assumption fails catastrophically, as illustrated in Figure 1.[7]

---

[7] https://www.nbcnews.com/meet-the-press/video/giuliani-truth-isn-t-truth-1302113347986

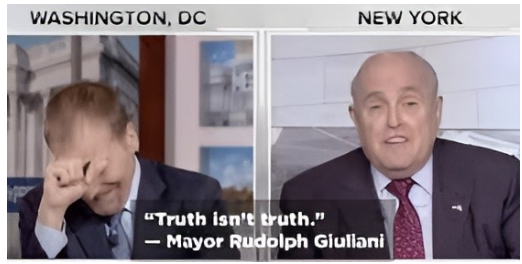

**Figure 1**: NBC interview hosted by **Chuck Todd**.

Apart from that, the next assumptions might, arguably, be considered more realistic:

- No sentence can, at the same time and under the same *aspect*, be both true and not true.
- Non monotonicity comes as a consequence of intentional or automatic fact updates; the rules themselves can be modified in law, even in natural sciences as happened after the advent of relativity and quantum mechanics; mathematical conjectures can either be disproved or turned into theorems.
- A domain is an 'open world', in which conclusions are highly context dependent. The expected result of an experiment may depend on assumed standard temperature and pressure conditions (STP), which, for instance, astronauts will not encounter outside Earth's atmosphere.
- Conflicting rules may coexist, sometimes with different weights, sometimes admitting diverging interpretations, and in law domains sometimes depending on conclusive ('iuris et de iure') or rebuttable ('iuris tantum') presumptions.[8]
- Operation execution can be non-deterministic: different effects or even failure may result; the execution may provoke unanticipated side effects; the context may undergo parallel modifications during execution which may modify the effects or even cause partial or total failure.
- Agents may have incomplete or false beliefs about the domain, concerning facts, operations, and the rules themselves; they can be the victims or the perpetrators of errors, lies, frauds, and may have wrong expectations about possible future worlds.
- Rational behavior can be frustrated by emotional decisions, and can be affected in multi-agent environments by goal/plan interferences.

About the last item above, it is worthwhile to cite the case of the English philosopher Herbert Spencer, supposedly a paragon of rationality, when he was wondering whether or not he should migrate to New Zealand [Durant]: "He made parallel lists of reasons for and against the move, giving each reason a numerical value. The sums being 110 points for remaining in England and 301 for going, he remained."

Clearly, being a philosopher, Spencer felt obliged to choose rationally – but, when faced with the consequences of the rational choice, took the opposite option, very likely motivated by emotive considerations. It is curious to notice how cleverly he, at first, was employing a *utility function* to compare the two plans. So $\Pi$ would in turn correspond to moving and not moving to New Zealand:

$$U(\Pi) = \sum_i p_i \times u_i, \text{ for } i = 1, 2, ..., n$$

Instead of utility functions we shall turn our attention to certain logic extensions, which we claim to be useful – even if still not sufficient – additions to first-order logic. By confining ourselves to logic we must alert the reader that we are not contributing to solve Spencer's dilemma, since emotions and personality traits will not be contemplated. The logic extensions will be gradually introduced, while we accompany the imaginary story of Joe, a student engaged in a university program.

At the first term of the program, Joe was aware that he should take exactly the set of courses prescribed to all new beginning students. So, at this point, he did not have to find out about any other course in the program. More generally, he got used to look exclusively for *information relevant at the given moment*. Some items of interest seemed obvious, there being no need to enquire. For instance, Joe assumed *by default* that if, in his opinion, he did well in an examination he would earn a high mark. Default expectations can by nature be frustrated since the examiner might disagree with some of Joe's results. Given a well-known puzzle, the 'missionaries and cannibals problem'[9], Joe asked the examiner if the possible existence of more than one boat, or of a nearby bridge, would affect the problem. To which the examiner replied that, even to that, Joe should employ the maxim of concentrating on the information declared relevant at the moment, which introduces one of our logic extensions: *circumscription*, recalling how this 'missionaries and cannibals problem' was employed as motivating example in [McCarthy].

---

[8] https://en.wikipedia.org/wiki/Presumption
[9] https://en.wikipedia.org/wiki/Missionaries_and_cannibals_problem

Having obtained a passing grade and preparing himself for the next term, Joe was told that now he was allowed to choose from a vast list of courses. Did he have to take all of them? The truth of the sentence 'student S has to take course C', Joe learned, was conditional. Course C could be either mandatory or elective, and it might even be restricted, only permitted if S had taken certain prerequisite courses. To us this qualification brings to mind **modal logic** [Blackburn et al.], more specifically associated with the *deontic*[10] aspect, which is included together with two other typical aspects in Table 1.

Table 1: Modal logics.

| Logic | Symbols | Expressions Symbolized |
|---|---|---|
| Modal Logic | □ | It is necessary that … |
|  | ◊ | It is possible that … |
| Deontic Logic | O | It is obligatory that … |
|  | P | It is permitted that … |
|  | F | It is forbidden that … |
| Temporal Logic | G | It will always be the case that … |
|  | F | It will be the case that … |
|  | H | It has always been the case that … |
|  | P | It was the case that … |
| Doxastic Logic | $B_x$ | x believes that … |

When the classes started again, Joe discovered that course c32, which he so much wished to take, had been suspended for not attaining the required minimum number of enrolments. Joe was sure that he had a right to take the course, since he had fulfilled all that was required in the program regulations. However, the same regulations imposed the minimum attendance required, and that had a stronger force than the rights of regularly enrolled students. To Joe, this was a disappointment, but to us it is an opportunity to illustrate the **defeasible reasoning** [Pollock] extension, whereby a rule can be superseded by a contrary more authoritative rule. Legal argument provides the most striking examples of rules defeating other rules. In a democratic country, law hierarchy principles place the constitution above federal laws, and federal laws above state laws. Indeed, legal reasoning has characteristics widely different from physics and mathematic, as convincingly argued by Stephen Toulmin [Toulmin], whose modified legal rule syllogism is shown in Figure 2. To antecedent, conclusion and rule, he adds the 'backing', which is the legal duly sanctioned provision (without which neither a law can be enforced, nor a judicial decision can be proffered), the 'rebuttal', which are circumstances that can defeat the rule application, and to the conclusion he adds a prefix expressing modality (the 'So, presumably' showed in Figure 2).

When he was deciding what courses to take in a later term, when he had already learned quite a lot about the subject, Joe took more time to choose the courses and also to find a way of earning some money in order to complement his meagre scholarship. That was not an easy task: he had to draw a schedule where there would be no timetable conflicts between the courses, and some time might be reserved for work. And, about working, he was warned that to keep his scholarship he could not be hired as a regular employee, but only as a trainee with a small token payment. For us, this effort to put together all activities in a conflict-free fashion can be supported by **argumentation theory** [Dung]. Indeed, this theory has been applied in the area of *requirements specification*, to conciliate the often conflicting objectives of the various stakeholders who jointly control a business enterprise, before effectively embarking on the design of an information system [Elrakaiby et al.].

---

[10] The term *deontic* is derived from the Ancient Greek: δέον, romanized: *déon* (gen.: δέοντος, *déontos*), meaning "that which is binding or proper." - https://en.wikipedia.org/wiki/Deontic_logic

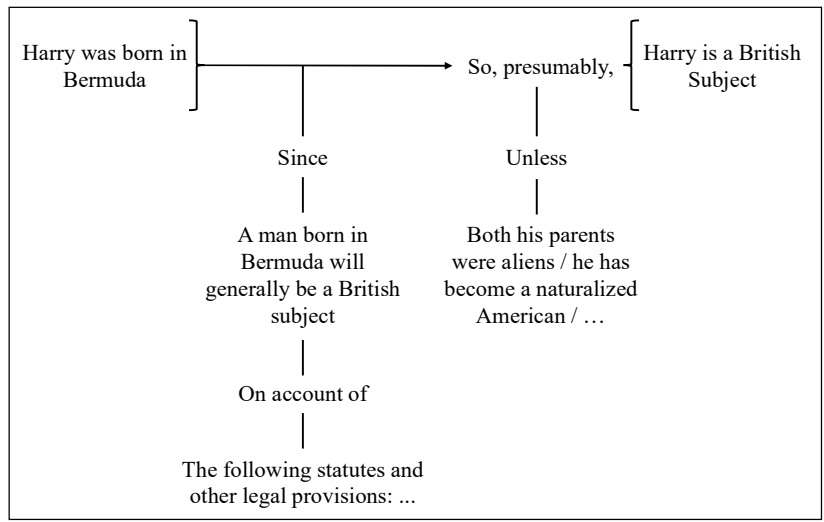

**Figure 2**: Toulmin's legal syllogism.

Surely, Joe wanted to complete his studies successfully and, with that in mind, tried to get acquainted with as many cases of success as possible. A friendly veteran, Moe, told him that those who managed to take course c32 no later than in the third term never failed to graduate. Other colleagues disagreed with Moe, claiming that a traineeship award was the decisive factor. Joe wrote both indications on a piece of paper:

- Taking course c32 → graduation
- Traineeship award → graduation

Joe's current schedule would prevent his adoption of both tactics. What he finally chose is not relevant to us. What is relevant is that Joe interpreted correctly the situation. Even assuming that both rules were verified with high frequency in the program's records, he understood that his choice would not proceed by deduction (i.e., given the antecedent of a rule conclude the consequent). Joe had to look at the rule in the inverse direction – if graduation is supposed to have occurred, then each antecedent should be taken as a *hypothesis*. Truly, Joe's reasoning was not fallacious. To us, this inverse of deduction has a name, **abduction** [Pereira and Saptawijaya]. It corresponds to what physicians and detectives do all the time. Doctors do not see a disease – they examine symptoms and must go through differential diagnoses if different diseases cause nearly the same symptoms. Likewise, a detective starts from the observed clues, from the vestiges left by the suspects, to identify who among them is the criminal.

One day, Joe was assigned to join a classroom debate, in a world history course, whose topic was the events that led to the second world war. The assignment required that the students should, prior to the debate, read certain book chapters and selected papers. Joe thought that he could pass without that and confined his preparation to a quick internet search. This decision proved to be not so clever, since students who did read the appointed texts obtained much better marks. This negative experience led Joe to recognize that, if he had read those texts, in particular chapter 12 of William Shirer's book [Shirer], he would have done much better. To us, Joe's belated realization is an instance of **counterfactual reasoning** [Starr]. Incidentally, Joe's internet search was not entirely fruitless, since it led him to discover a link to a video[11] based on Shirer's chapter, produced with the help of ChatGPT, exploring a positive counterfactual response to the 'what if' question: "could World War II have been avoided if Chamberlain and Daladier had not yielded to Hitler's demands in Munich?"

The classroom debate also taught Joe unexpected lessons on human behavior. The participating students came from different nationalities and, in addition, several of them were members of exclusive groups, consistently obedient to the orientation of some outspoken leader. Some students were not even aware of all relevant facts in the historical incidents, which led to mistaken opinions. Opinions also varied with the passage of time in the course of the debate, as a student heard what the others had said, and frames of mind were strongly affected by the heated battle of viewpoints. Joe learned then that factual knowledge tended in practice to be superseded by biased opinions, as we are told when studying **belief logic** [Fagin and Halpern].

---

[11] https://narrativelab.org/chatgeppetto/video.php?id=104

To act in a rational way, Joe tried to organize his life, both with respect to his studies and to his future professional career, in terms of the objectives he wanted to fulfill – in other words, in view of the *possible future states* that would result from his actions. He sought to have a precise idea about the allowed actions, characterizing each one by its precondition and probable effects. He then formulated a sequence of actions, committing himself to their execution, even knowing beforehand that circumstances could vary along the time that might interfere with the result of the actions, and, even more critically, that he was relying on the favorable behavior of other people [Wilensky]. We would explain to Joe that what he was doing was to perform a *planning process*, for which we would recommend to have a look at **situation calculus** [Kowalski]. For a change, the basic notion behind this formal discipline is quite intuitive, and can be expressed by the following three second-order logic rules:

- A fact F holds if it is true at the initial state.
- F holds if it is <u>added</u> as one of the effects of an operation Op, and the <u>preconditions</u> of Op hold at the current state.
- F holds after the execution of an operation Op if it did already hold at the current state and if it is not <u>deleted</u> as one of the effects of Op.

the third rule being included to handle the *frame problem*, which is the need, unavoidable in first-order logic formulation, to determine all facts that would *not* be affected by the execution of operation Op.

We are now in a position to sketch, still in informal terms, a roadmap towards a *composite logic strategy* to specify a domain as an elementary component of a multiverse project:

1. Consider adopting other forms of reasoning, such as
    - Defeasible reasoning: Compare the strength of old/new defeaters and warrants.
    - Argumentation theory: Put together or revise overall compatible interpretations.
    - Abductive reasoning: Formulate hypotheses on what may cause certain situations of interest.
2. Consider adopting non-monotonic reasoning, such as
    - Circumscription: Delimit the initial context of interest.
3. Consider adopting a "possible worlds" form of reasoning, such as:
    - Counterfactual logic: Explore the possible worlds that can emerge.
    - Belief logic: Introduce the points of view of individual agents and groups (In doxastic logic, belief is treated as a modal operator).
4. Consider adopting a formalism for reasoning about dynamical domains, such as:
    - Situation calculus: Find how agents can schedule actions to achieve objectives.

A composite logic, as the name implies, would combine one or more formal systems from each of these axes to provide a framework to reason about the domain of interest.

Modal logics, though not explicitly included in the above list, would seem to be a pervasive component of any combination, serving to determine rule applicability with respect to diverse *aspects*. The work of [Fagin and Halpern], for instance, combines belief and time, recalling that belief and time are associated, respectively, with *doxastic* and *temporal* logic as forms of modal logic in table 1. In [Moses and Shoham] belief is treated as defeasible knowledge. To give yet another example, in our work on the use of interactive planning for plot composition [Ciarlini et al.], we combined situation calculus with temporal logic, thus allowing to express, in a swords-and-dragons world, goal inference rules such as: "If a villain kidnaps a victim, the victim will be *eventually* rescued".

An attempt at a logic combination that seems somewhat close to our proposed strategy is John Sowa's modelling approach. The author begins by deploring inadequate views of possible worlds [Sowa]:

> Since the pioneering work by Kripke and Montague, the term possible world has appeared in most theories of formal semantics for modal logics, natural languages, and knowledge-based systems. Yet that term obscures many questions about the relationships between the real world, various models of the world, and descriptions of those models in either formal languages or natural languages. Each step in that progression is an abstraction from the overwhelming complexity of the world. At the end, nothing is left but a colorful metaphor for an undefined element of a set **W** called worlds, which are related by an undefined and undefinable primitive relation **R** called accessibility.

He ends up with a combination different from ours but equally aiming at some meaningful modelling theory of possible worlds, in terms of agents, actions, planning, laws and facts, noting one vital point in common with our composite strategy: the appeal to *multimodal reasoning* [Sowa]:

> As this article has shown, semiotics goes beyond relationships among signs to relationships of signs to the world and to the *agents* who observe the world, interpret it in signs, and use the signs to *plan further actions* upon the world. For artificial intelligence, this connection of signs to perception, action, and language supports the symbol grounding necessary for making the language processing by AI systems and the actions by AI robots meaningful in the same sense that the words and actions of humans are meaningful. The combination of semiotics with Dunn's semantics of laws and facts provides a theoretical foundation for modality and intentionality that captures more of the intended interpretation than a undefinable relation **R** over an undefined set **W**. An important promise of this *combination* is the ability to support *multimodal reasoning* as a kind of metalevel reasoning about the source of the *laws* and *facts*. Instead of complex axioms for each mode with even more complex interactions between modes, it enables the laws to be partitioned in a hierarchy that represents grades of necessity or levels of entrenchment: logical, physical, economic, legal, social, cultural, or personal.

Starting to figure how to move towards the design of a module to implement our proposed composite logic strategy in a metaverse architecture, we thought that incorporating an AI agent, such as ChatGPT, would bring in its extensive world knowledge and would allow to set up a natural language interface able to hide from users the complex axioms needed to ensure the consistency of the combinations. With that in mind, we did an experiment with ChatGPT, in order to discover whether it had acquired some basic knowledge about the logic extensions as part of its training. The result of the experiment is reported in the next section.

## 3. The logical extensions as viewed by ChatGPT

We owe this section to ChatGPT, to which we asked to reply whether it was aware of each of the logical extensions treated in this study, and, if the response was positive, we asked for an explanatory example. Our few insertions are enclosed in angular brackets.

The results are encouraging with respect to all extensions, as will be noted in the respective sub-sections. One caveat, however, is in order: having a basic knowledge of the rules of a theory does not imply that ChatGPT would be able to correctly execute an algorithm that applies such rules.

For instance, ChatGPT had no difficulty with a simple plan to move a block from position *p1* to position *p2*. But it failed to generate and execute a plan in the quite elementary 'monkey-and-bananas problem',[12] even being given a complete set of rules transcribed from a Prolog program. In fairness, we must point out that the creators of AI agents, early realizing this lack, provided features to achieve a two-way access to algorithms developed in a variety of programming languages, as will be discussed in Section 4.

### 3.1. Circumscription

Circumscription is a non-monotonic logic created by John McCarthy. It is a form of default reasoning that allows for making assumptions and drawing conclusions in the absence of complete information.

One example of the application of circumscription is in the field of artificial intelligence and commonsense reasoning. Circumscription is used to reason about default assumptions and exceptions in a knowledge representation system.

For instance, consider the following scenario:

Assumption:
1. Birds usually fly.

With circumscription, we can represent this assumption as a default rule:

$Birds(X) \rightarrow Fly(X)$

---

[12] https://en.wikipedia.org/wiki/Monkey_and_banana_problem

This rule expresses that it is generally true that birds fly. However, circumscription also allows for exceptions to be specified explicitly. For example, we can introduce an exception:

Birds(penguin) → ¬Fly(penguin)

This exception states that penguins, despite being birds, do not fly.

Now, let's consider a specific case:

Birds(tweety)

Based on the default rule and the exception, circumscription allows us to draw conclusions even when the complete information is not available. In this case, we can conclude:

Fly(tweety)

Here, circumscription enables us to make a default assumption based on the general rule that birds usually fly. However, if we had more specific information about tweety being a penguin, we would have concluded that tweety does not fly, due to the exception.

Circumscription provides a way to reason about defaults and exceptions in a non-monotonic manner, allowing for reasoning in situations where complete information may be lacking or when we want to capture typical behavior while allowing for specific exceptions. It has applications in various domains, such as knowledge representation, planning, and commonsense reasoning.

## 3.2. Modal logic

Modal logic is a branch of formal logic that deals with the concepts of necessity and possibility. It extends classical logic by introducing modal operators such as "necessarily" and "possibly" to express statements about what is necessarily true or possible.

An example of modal logic's application is in reasoning about knowledge and belief. In epistemic logic, a branch of modal logic, we use modal operators to reason about what agents know or believe. For instance, consider the following statement:

"Agent A knows that it is raining."

In modal logic, we can represent this statement using a modal operator, typically denoted as "$K$," to signify knowledge. The statement can be represented as:

"$K$A: It is raining."

This formulation allows us to reason about knowledge explicitly. We can analyze statements like "If A knows that P, then P must be true" or "If A knows that P, then A believes that P." Modal logic provides a formal framework for reasoning about knowledge and belief, enabling us to make precise statements and draw logical conclusions about these concepts.

Epistemic logic is just one example of modal logic's application. Modal logic is also used in other areas such as philosophy, computer science, linguistics, and artificial intelligence to reason about necessity, possibility, time, action, and other modalities.

## 3.3. Defeasible reasoning

Defeasible reasoning is a form of reasoning that deals with arguments and reasoning patterns that are presumptive and subject to exceptions or rebuttals. It allows for reasoning based on general rules or assumptions that can be overridden or defeated by specific conditions or exceptions.

One example of defeasible reasoning is found in legal reasoning. In legal contexts, there are general rules or presumptions that apply to certain situations, but these rules can be defeated or overridden by specific circumstances. For instance, consider the following argument:

Argument:
1. John is the owner of the property.
2. The owner has the right to access the property.
3. Therefore, John has the right to access the property.

In this argument, we have a general rule that states the owner has the right to access the property. However, this rule can be defeated or overridden if there is evidence or a specific circumstance that suggests otherwise. For example, if there is a legal agreement or court order that restricts John's access to the property, then the general rule is defeated, and John's right to access the property is negated.

Defeasible reasoning allows us to reason in a way that takes into account exceptions, rebuttals, and conflicting evidence. It acknowledges that certain rules or assumptions may hold most of the time but can be overridden or defeated in specific cases. This form of reasoning is particularly useful in situations where there is uncertainty, incomplete information, or conflicting evidence, as it allows for more flexible and context-sensitive reasoning.

**3.4. Argumentation theory**

<*This is about*> the theory of argumentation as presented by Phan Minh Dung in "On the Acceptability of Arguments and its Fundamental Role in Nonmonotonic Reasoning, Logic Programming and N-Person Games" (Artificial Intelligence, 1995). Dung's argumentation framework, often referred to as Dung's AF, is a formal model for representing and evaluating arguments and their relationships.

In Dung's argumentation framework, arguments are represented as nodes, and relationships between arguments are represented as attacks. An argument attacks another argument when it provides a reason to doubt or refute it. The acceptability of arguments is determined by the concept of admissibility, which considers the presence or absence of cycles in the attack relationships.

Let's consider a simple example to demonstrate the application of Dung's argumentation framework:

Arguments:
- A: It is raining outside.
- B: I forgot my umbrella.
- C: I should take a taxi.

Attacks:
- A attacks B: If it is raining outside, you should have remembered your umbrella.
- B attacks C: If I forgot my umbrella, I should take a taxi.

In this example, argument A states that it is raining outside, while argument B claims that the person forgot their umbrella. Argument C suggests taking a taxi as a solution. Argument A attacks argument B because if it is indeed raining outside, the person should have remembered their umbrella. Argument B, in turn, attacks argument C by asserting that if the person forgot their umbrella, taking a taxi would be a suitable alternative.

To determine the admissibility of arguments, we need to check for cycles in the attack relationships. In this case, we have a cycle between A, B, and C. A attacks B, B attacks C, and C attacks A. This cycle indicates a potential inconsistency or conflict in the arguments.

If we were to apply Dung's AF to evaluate the arguments in this example, the presence of a cycle would render the set of arguments as not admissible. In other words, there is an inconsistency within the arguments, and no definitive conclusion can be drawn.

Dung's argumentation framework provides a formal basis for analyzing and evaluating arguments in a logical and systematic manner, allowing us to assess the acceptability and consistency of arguments within a given context.

**3.5. Abduction**

<*This is about*> the concept of abduction as introduced by Charles Sanders Peirce. Abduction, in the context of Peirce's work on the logic of science, refers to a form of reasoning that involves generating explanatory hypotheses to account for observed phenomena or facts. It is often described as "inference to the best explanation".

In abduction, an inference is made from an observed effect or set of facts to the most plausible cause or explanation that could account for those observations. The process of abduction involves proposing a hypothesis that, if true, would provide the best explanation for the observed evidence. This hypothesis is then tested and refined through further observation and experimentation.

Here's an example to illustrate the application of abduction:

Let's say you wake up one morning and find your garden has been partially destroyed. Some plants are uprooted, and there are large footprints in the soil. You observe that the fence around your garden is damaged in one area. Using abductive reasoning, you generate several possible explanations for the observed evidence:

1. Explanation 1 (Hypothesis): A strong windstorm passed through the area, causing the plants to be uprooted and damaging the fence.
2. Explanation 2 (Hypothesis): An animal, such as a deer or a bear, entered the garden, uprooted the plants, and damaged the fence.
3. Explanation 3 (Hypothesis): A person trespassed into your garden, deliberately damaging the plants and fence.

In this scenario, abduction involves considering various hypotheses that could explain the observed phenomena. Each hypothesis provides a potential explanation, but further investigation and evidence gathering would be needed to determine the most likely cause. Through additional observations, such as checking for tracks or speaking with neighbors, you might gather more evidence to support or refute each hypothesis, eventually leading to a more informed conclusion.

Abduction is a valuable reasoning process in scientific inquiry and problem-solving, as it helps generate initial explanations or hypotheses that can be further investigated and tested.

## 3.6. Counterfactual logic

Counterfactual reasoning involves considering hypothetical or contrary-to-fact scenarios and reasoning about the outcomes that would have occurred if certain conditions or events were different. It involves imagining what could have happened if the past had been different.

One example of a counterfactual rule is the "but for" test. It is commonly used in legal contexts to determine causation. The rule states that if an event A is claimed to have caused event B, then A is considered the cause of B if, but for the occurrence of A, B would not have occurred. In other words, if we can establish that event B would not have happened if event A had not occurred, then event A is considered the cause of event B.

For example, let's say there was a car accident, and the driver of Car A rear-ended Car B. Car B's driver suffered injuries and filed a lawsuit claiming that Car A's driver caused the accident. To determine causation using the counterfactual rule, we would consider whether the accident would have occurred if Car A's driver had not rear-ended Car B. If it can be established that the accident would not have happened "but for" the rear-ending by Car A, then Car A's driver would be considered the cause of the accident according to the counterfactual rule.

<*Authors' question: Now consider the following case: "John did not study for the Math examination, and failed to pass in Math." Please formulate an "if" counterfactual rule explicitly referring to the specific better result that John might have obtained.*>

In the given case, where John did not study for the Math examination and failed to pass, an "if" counterfactual rule could be formulated as follows:

"If John had studied for the Math examination, he might have obtained a better result and passed the exam."

This counterfactual statement reflects the hypothetical scenario where John did study for the Math examination, leading to a different outcome, specifically a better result and successfully passing the exam.

## 3.7. Belief logic

Belief logic, also known as doxastic logic, is a branch of modal logic that focuses on reasoning about beliefs and knowledge. It provides a formal framework for analyzing and reasoning about agents' beliefs, their relationships, and their dynamics.

One example of the application of belief logic is in the field of multi-agent systems and artificial intelligence. In multi-agent systems, agents interact and reason about the beliefs and knowledge of other agents. Belief logic helps in representing and reasoning about these beliefs and their dynamics.

For instance, consider a scenario with two agents, Alice and Bob. They have a shared goal of meeting at a specific location but have limited information about each other's whereabouts. Using belief logic, we can model their beliefs and reason about their actions.

Let's define some belief-related propositions:

Beliefs:
- Believes(Alice, p): Alice believes that proposition p.
- Believes(Bob, q): Bob believes that proposition q.

We can then use the belief logic operators to reason about their beliefs:

Example Belief Logic Statement:
- Believes(Alice, MeetAt(LocationX)) → Believes(Bob, MeetAt(LocationX))

This statement represents that if Alice believes that they should meet at LocationX, then Bob believes the same.

With belief logic, we can reason about the consequences of these beliefs, evaluate their consistency or inconsistency, and analyze the impact of belief revisions or updates. For example, if Alice receives new information that Bob is not available at LocationX, her belief might change, and we can reason about the effect of this belief update on Bob's beliefs and subsequent actions.

Belief logic is used in various fields, including multi-agent systems, distributed computing, game theory, and decision theory, to model and reason about agents' beliefs, knowledge, and their interactions. It provides a formal framework for studying belief dynamics, belief revision, and other aspects of agents' epistemic states.

## 3.8. Situation calculus

Situation Calculus is a logical formalism developed by John McCarthy for representing and reasoning about actions, change, and dynamic aspects of the world.

One example of the application of Situation Calculus is in the field of automated planning. Automated planning involves generating sequences of actions to achieve desired goals in a dynamic environment. The Situation Calculus provides a framework to represent and reason about the effects of actions and how the world evolves over time.

Consider a planning problem where we want to model a robot navigating a grid world. The robot can move in four directions: up, down, left, and right. We can represent the actions and their effects using the Situation Calculus.

Let's define some predicates and action descriptions:

Predicates:
- At(x, y): The robot is at position (x, y).
- Clear(x, y): The position (x, y) is clear of obstacles.

Action descriptions:
- MoveUp: Causes the robot to move up if the position above is clear.
- MoveDown: Causes the robot to move down if the position below is clear.
- MoveLeft: Causes the robot to move left if the position to the left is clear.
- MoveRight: Causes the robot to move right if the position to the right is clear.

Using the Situation Calculus, we can define the effects of these actions and how the world changes:

Effects:
- MoveUp: At(x, y) ∧ Clear(x, y+1) → At(x, y+1)
- MoveDown: At(x, y) ∧ Clear(x, y-1) → At(x, y-1)
- MoveLeft: At(x, y) ∧ Clear(x-1, y) → At(x-1, y)
- MoveRight: At(x, y) ∧ Clear(x+1, y) → At(x+1, y)

By representing the actions and their effects in the Situation Calculus, we can reason about the robot's possible actions and how they impact the robot's position. We can formulate planning problems by specifying initial states, goals, and constraints, and then use automated planning algorithms to generate action sequences that achieve the desired goals while respecting the given constraints.

The Situation Calculus provides a formal framework to reason about actions, change, and dynamic aspects of the world, making it a valuable tool for modeling and solving planning problems in various domains, including robotics, logistics, and automation.

## 4. First steps towards a two-way connection between ChatGPT and Prolog

As demonstrated in the previous section, ChatGPT has the basic knowledge to explain the logic extensions explored in this work, but it is not able to perform the complex reasoning processes required to correctly apply some of these logic extensions. To provide a simple example, we attempted to make ChatGPT produce a plan of actions for the classic monkey-and-bananas problem using situation calculus.

We provided to ChatGPT a complete description of the problem and the actions available for the monkey:

*<Consider the following facts: The Monkey is at the Door. The Bananas are at the Center. The Box is at the Window. The Monkey and the Box have height Low. The Bananas have height High.*

*Also consider the following actions that are available for the Monkey:*

*Action: Go, which moves the Monkey from one location to a new location;*
*Preconditions: the Monkey must have height Low;*
*Effects: the Monkey will be at the new location.*

*Action: Push, which moves an object from one location to a new location;*
*Preconditions: both Monkey and object must have height Low and be at the same location;*
*Effects: both Monkey and object will be at the new location.*

*Action: ClimbUp, which allows the Monkey to climb onto an object;*
*Preconditions: both Monkey and object must have height Low and must be at the same location;*
*Effects: the Monkey will have height High.*

*Action: ClimbDown, which allows the Monkey to climb down from an object;*
*Preconditions: the Monkey must have height High and both the Monkey and the object must be at the same location;*
*Effects: the Monkey will have height Low.*

*Action: Get, which allows the Monkey to get and hold an object;*
*Preconditions: both Monkey and object must have height High and be at the same location;*
*Effects: the Monkey will be holding the object.*

*Based on the previous definitions, write a sequence of actions for the Monkey to be holding the Bananas.>*

The results are varied and inconsistent, but most of the times ChatGPT provides incorrect plans, such as:

To have the Monkey hold the Bananas, we can follow these steps:

1. Go to the Door.
2. ClimbUp onto the Box (now the Monkey has height High).
3. Push the Bananas from the Center to the Door.
4. ClimbDown from the Box (now the Monkey has height Low again).

5. Get the Bananas (now the Monkey is holding the Bananas).

So the sequence of actions is:

1. Go to the Door.
2. ClimbUp onto the Box.
3. Push the Bananas from the Center to the Door.
4. ClimbDown from the Box.
5. Get the Bananas.

Given the limited logic reasoning capabilities of ChatGPT, we started to explore ways for achieving a two-way integration between Prolog and ChatGPT, with the former acting as the logic engine that provides reasoning functionalities to ChatGPT, and the latter offering natural language generation capabilities to Prolog.

**4.1. Providing Prolog functionalities to ChatGPT**

Considering the difficulty that ChatGPT has in solving the simple monkey-and-bananas problem, our initial experiment to evaluate the possibilities of using Prolog as a logic engine consisted in providing to ChatGPT information about a function that can calculate the monkey's plan for a given initial state. Our goal is to allow ChatGPT to use its natural language processing capabilities to analyze the user input and identify when to call the function we provided in order to obtain the requested plan. In this way, our program can identify when ChatGPT wants to call Prolog, execute the requested function, and provide the resulting plan back to ChatGPT. The implementation of this process is possible thanks to the function calling functionality recently added to the OpenAI API.[13]

The complete code of the program implemented to test the integration between ChatGPT and Prolog is presented in Appendix A. The program consists of two code files: (1) a Prolog file called *monkey_and_bananas.pl*, which implements a predicate called *plan* that can be used to find a sequence of actions to solve the monkey-and-bananas problem; and (2) a Python script called *monkey_and_bananas.py*, which implements the process of running the Prolog environment through the swiplserver library,[14] and also manages the communication between ChatGPT and the Prolog environment. An example of the results produced by the program is:

**python monkey-and-bananas.py**

**> Enter a prompt:** <*Write a sequence of actions that lead the monkey to get the bananas considering that he starts at the door, he is on the ground, the box is at the window, and the monkey does not have the banana.*>

**> ChatGPT asked to run function:** get_monkey_plan({'monkey_start_ground_location': 'at_door', 'monkey_start_height_location': 'on_ground', 'box_start_location': 'at_window', 'monkey_has_banana': 'no_banana'})

**> Plan generated by Prolog and sent to ChatGPT:** [{"PLAN": [{"args": ["at_door", "at_window"], "functor": "walk"}, {"args": ["at_window", "at_center"], "functor": "push_box"}, "climb_box", "get_banana"]}]

**> Response from ChatGPT:**
Here is a sequence of actions for the monkey to get the bananas:
1. Walk from the door to the window.
2. Push the box from the window to the center of the room.
3. Climb on top of the box.
4. Reach out and grab the bananas.
By following these actions, the monkey will successfully obtain the bananas.

It is also interesting to notice that ChatGPT is responsible for deciding when to call the Prolog function. If answering the user prompt is possible without relying on the Prolog function, ChatGPT uses its own knowledge to formulate a response:

---

[13] https://platform.openai.com/docs/guides/gpt/function-calling
[14] https://www.swi-prolog.org/packages/mqi/prologmqi.html

**python monkey-and-bananas.py**

**> Enter a prompt:** <*What is Prolog?*>

**> Response from ChatGPT:**
Prolog is a logic programming language that is based on first-order logic. It was developed in the early 1970s and is widely used in areas such as artificial intelligence, natural language processing, and expert systems. In Prolog, programs are expressed as a set of logical rules and facts, and computations are performed by applying logical inference. Prolog is particularly well-suited for problems that can be described in terms of relations and unification. It supports pattern matching, backtracking, and automatic handling of logical variables.

**4.2. Accessing ChatGPT functionalities from Prolog**

While allowing ChatGPT to access Prolog predicates can provide accurate logical calculations, the other direction of communication is also worth investigating. The excellent natural language generation compatibilities of ChatGPT can also be used to benefit Prolog applications. Although we cannot access the OpenAI API[15] directly from Prolog yet, it is possible to implement a custom API to act as a proxy between Prolog and ChatGPT.

In order to explore this interaction, we implemented the two programs presented in Appendix B. The first is a Python API script called gpt-prolog.py, which runs on a Web server and receives Web requests from the Prolog environment. The requests consist of text prompts that are forwarded to ChatGPT through the OpenAI API. When the Python script gets a response from ChatGPT, the response is sent to the Prolog environment. The second program is a Prolog implementation that includes a predicate to send requests to the Python API script (*ask_chatgpt*) and two special predicates, *narrate_story* and *narrate_events*, which were designed to provide, when called respectively by test1 and test2, either a narrative of the entire story or separate event-by-event narratives.

Considering the following sequence of events:

ride('Draco', 'White Castle'), overpower('Draco','Walt'), capture('Draco','Princess Marian'), carry('Draco','Princess Marian','Fortress of Draco'), warn('Walt','Sir Brian','capture'), ride('Sir Brian','Fortress of Draco'), fight('Sir Brian','Draco'), defeat('Sir Brian','Draco'), free('Sir Brian','Princess Marian'), carry('Sir Brian','Princess Marian','White Castle')

The Prolog program can produce, with the assistance of ChatGPT, the following narratives:

**?- test1(Lst).**

Lst = [
'Draco, a notorious villain, rode his horse towards the White Castle with the intention of causing trouble. The sentinel of the castle, Walt, tried to overpower Draco, but the villain proved to be too strong and was able to capture Princess Marian, a resident of the castle. Draco then took the princess and carried her to the Fortress of Draco.',

'Walt knew that he could not handle the situation on his own, so he warned Sir Brian, a brave hero, about the crime that was committed. The hero rode his horse towards the Fortress of Draco to face the villain.',

'Upon reaching the fortress, Sir Brian engaged in a fierce fight against Draco. In the end, the hero emerged victorious and defeated the villain. He then freed Princess Marian from captivity and carried her back to the safety of the White Castle.'
]

**?- test2(Lst).**

Lst = [
'Draco rides towards the White Castle. ',
'Walt, the sentinel, tries to stop him, but Draco overpowers him.',
'Draco captures Princess Marian.',
'Draco carries Princess Marian to his Fortress.',
'Walt warns Sir Brian of the capture.',

---

[15] https://platform.openai.com/docs/api-reference

```
        'Sir Brian rides to the Fortress of Draco.',
        'Sir Brian fights Draco.',
        'Sir Brian defeats Draco.',
        'Sir Brian frees Princess Marian.',
        'Sir Brian carries Princess Marian to the White Castle.'
    ]
```

The results produced in this experiment far surpass the quality and variety of the textual narrations we were able to obtain in many of our previous works in the interactive storytelling domain using templates to transform narrative events into text [Silva et al.; Lima et al. 1; Barbosa et al.; Lima et al. 2; Lima et al. 3].

## 5. Concluding remarks

The logic extensions of our interest were discussed in a rather limited way in this study, simply at the application domain level. Future work is needed, not only to achieve a more formal presentation[16], but mainly to make a transition from single application domains to universes containing such domains, and then proceeding to attain a multiverse/metaverse scope.

We mentioned that plans can lead from initial states to *possible future states*. It would be nice to define operators leading from a firstly specified domain into other domains in a given universe. The possibly many initial domains could then be regarded as centroids from which would radiate the domains – forming separate or intersecting clusters – obtained by applying such domain-transformation operators. Previous works of our group [Magalhães et al.; Sacramento et al.] are likely to help future advances in this direction.

Even more generally, we expect that, combining mathematical logic with the natural language understanding/generation ability and extensive fact and expertise repository afforded by AI agents, such as ChatGPT – enhanced by bidirectional access to algorithms developed in suitable programming languages – should be instrumental to guide metaverse projects.

On the opposite direction, there is currently a movement[17] that raises an alert against the dangers of the ever-growing progress of artificial intelligence. The fear that artificial intelligence could in a not distant future take the form of sentient human-like robots, emerges in the following dialog between two very advanced positronic robots, George Nine and George Ten, in Isaac Asimov's tale *That Thou Art Mindful of Him*[18] [Asimov 1]:

> [George Ten:] "Of the reasoning individuals you have met, who possesses the mind, character and knowledge that you find superior to the rest, disregarding shape and form, since that is irrelevant?"
> "You," whispered George Nine.
> "But I am a robot. There is in your brain paths a criterion for distinguishing between a robot of metal and a human being of flesh.
> How then can you classify me as a human being?"
> "Because there is in my brain paths an urgent need to disregard shape and form in judging human beings, and it rises superior to the distinction between metal and flesh. You are a human being, George Ten, and more fit than the others."
> "And I find that of you," whispered George Ten. "By the criteria of judgment built into ourselves, then, *we find ourselves to be human beings within the meaning of the Three Laws, and human beings, moreover to be given priority over those others*."

Fortunately there are limits – for the time being… – to the achievements of AI agents. In Section 4 we mentioned that, for developing and executing algorithms of some complexity, it was found necessary to permit access to and from suitable programming languages, such as Python and Prolog, but a stronger argument can be alleged: AI agents still cannot match the creative power of talented humans. We had a proof of that when consulting ChatGPT about how to intelligently avoid a social/political blunder, narrated by Voltaire in his *Zadig* novel [Voltaire]. Our dialog with ChatGPT went as follows:

> <Consider the following narrative: "There had continued at Babylon, for the space of fifteen hundred years, a violent contest that had divided the empire into two sects. The one pretended that they ought to enter the temple of Mithra with the left foot foremost; the other held this custom in detestation, and always entered with the right foot first." Please indicate how a person could enter that temple without displeasing one of those two sects.>

---

[16] Possibly adapting Kripke-style structures [Kripke].
[17] https://futureoflife.org/open-letter/pause-giant-ai-experiments/
[18] Taken from the Bible - Psalm 8:4-9 'What is man, that thou art mindful of him?'

ChatGPT: In order to enter the temple of Mithra in Babylon without displeasing either of the two sects, one could employ a strategy that promotes unity and respects the beliefs of both groups. Here's a possible approach:

….

3 . Enter the temple by taking a small step forward, while simultaneously extending the left foot backward in a manner that is not prominent or exaggerated.

....

<We then asked ChatGPT: Do you know how Zadig managed to enter the temple without displeasing either of the two sects?>

ChatGPT: In Voltaire's novel "Zadig," the titular character faces a dilemma regarding the custom of entering the temple of Mithra in Babylon. One sect believed in entering with the left foot first, while the other sect insisted on entering with the right foot first. The people were curious to see which sect Zadig would support during the celebration of the sacred fire.
Zadig ingeniously resolves the situation by devising a solution that appeases both sects. Instead of entering the temple with either foot first, he decides to leap into the temple with both feet together. This action allows Zadig to maintain a sense of neutrality and avoid favoring one sect over the other. By choosing a completely different approach, Zadig manages to uphold peace and harmony by not aligning himself with either faction's belief.

Clearly, ChatGPT's solution to Zadig's dilemma – 'enter the temple by taking a small step forward, while simultaneously extending the left foot backward' – was not adequate. The 'small step', with the right foot at a short distance from the left foot, would still privilege one of the two opposing groups. We had to remind ChatGPT to look at the clever solution found by Voltaire, and only then the AI agent hit upon it: 'to leap into the temple with both feet together'.

Deservedly famous writers can go much beyond this ability to solve puzzles. The farfetched notion of level IV multiverses [Tegmark], mentioned in Section 1, postulating universes governed by natural laws different from ours, was rendered plausible by the imaginative talent of the same science fiction author, Isaac Asimov, in his novel *The Gods Themselves* [Asimov 2]. The intelligent beings in a parallel universe, in which strong nuclear interaction was much stronger than here, taught our scientists a process whereby energy was abundantly generated in both universes – but not actually for our benefit, since its continued use would eventually cause the sun to explode. A brilliant, inquisitive scientist perceived this in time, and provided an elegant solution, based on the principle that, if two universes existed in parallel, then why not an infinite number? Working with the help of an 'intuitionist' female assistant, he contrived a channel to a third universe whose physical laws diverged in the opposite direction, thus managing to reestablish the balance and save Earth from imminent disaster.

Flesh-and-blood human beings still have a chance to play a role somewhere in the metaverse.

# Appendix A: Example of program that allows a Prolog predicate to be called by ChatGPT

Prolog Program (monkey_and_bananas.pl)

```
action(
   state(at_center, on_box, at_center, no_banana),
   get_banana,
   state(at_center, on_box, at_center, has_banana)
).
action(
   state(P, on_ground, P, B),
   climb_box,
   state(P, on_box, P, B)
).
action(
   state(P1, on_ground, P1, B),
   push_box(P1, P2),
   state(P2, on_ground, P2, B)
).
action(
   state(P1, on_ground, C, B),
   walk(P1, P2),
   state(P2, on_ground, C, B)
).

plan(state(_,_,_,has_banana),[]).
plan(S1,[M|R]) :-  action(S1,M,S2), plan(S2,R).
```

Python Program (monkey_and_bananas.py)

```
from swiplserver import PrologMQI
import openai
import json

openai.organization = OPENAI_ORG
openai.api_key = OPENAI_KEY

def calculate_plan(m_loc, m_height, b_loc, m_has_banana):
  with PrologMQI() as mqi:
    with mqi.create_thread() as prolog_thread:
      prolog_thread.query("set_prolog_flag(encoding, utf8).")
      prolog_thread.query("consult(\"monkey_and_bananas.pl\").")
      prolog_thread.query_async("plan(state(" + m_loc + ", " + m_height + ", " +
                      b_loc + ", " + m_has_banana + "), PLAN).", find_all=False)
      result = prolog_thread.query_async_result()
  return json.dumps(result)

def ask_chatgpt(prompt):
  response = openai.ChatCompletion.create(
    model ="gpt-3.5-turbo-0613",
    messages=[{"role": "user", "content": prompt}],
    functions=[{
      "name": "get_monkey_plan",
      "description": "Gets the sequence of actions that lead the monkey to get
                      the banana starting from a given state",
      "parameters": {
        "type": "object",
        "properties": {
          "monkey_start_ground_location": {"type": "string", "enum":
                                          ["at_center", "at_window", "at_door"]},
          "monkey_start_height_location": {"type": "string", "enum":
                                          ["on_ground", "on_box"]},
```

```python
                "box_start_location": {"type": "string", "enum":
                                    ["at_center", "at_window", "at_door"]},
                "monkey_has_banana": {"type": "string", "enum":
                                    ["has_banana", "no_banana"]},
            },
            "required":["monkey_start_ground_location",
                        "monkey_start_height_location",
                        "box_start_location",
                        "monkey_has_banana"],
        },
      }],
      function_call="auto",
    )
    message = response["choices"][0]["message"]
    if message.get("function_call"):
        function_name = message["function_call"]["name"]
        function_args = json.loads(message["function_call"]["arguments"])
        print("> ChatGPT asked to run function:" + function_name + "(" +
              str(function_args) + ")")
        function_response = calculate_plan(
          m_loc = function_args.get("monkey_start_ground_location"),
          m_height = function_args.get("monkey_start_height_location"),
          b_loc = function_args.get("box_start_location"),
          m_has_banana = function_args.get("monkey_has_banana")
        )
        print("> Plan generated by Prolog and sent to ChatGPT:" + function_response)
        second_response = openai.ChatCompletion.create(
          model="gpt-3.5-turbo-0613",
          messages=[
            {"role": "user", "content": prompt},
            message,
            {
              "role": "function",
              "name": function_name,
              "content": function_response,
            },
          ],
        )
        return second_response["choices"][0]["message"]["content"].strip()
    else:
        return message["content"].strip()

user_prompt = input()
plan = ask_chatgpt(user_prompt)
print("> Response from ChatGPT:\n" + plan)
```

# Appendix B: Example of plot-to-story transformation, calling ChatGPT from a Prolog environment

## Python Program (gpt-prolog.py)

```python
import openai
import cgi
import urllib

openai.organization = OPENAI_ORG
openai.api_key = OPENAI_KEY

def send_openai_request(messages):
  response = openai.ChatCompletion.create(
                      model="gpt-3.5-turbo",
                      messages=messages, temperature=.9,
                      max_tokens=2048, top_p=1,
                      frequency_penalty=0, presence_penalty=0)
  return = response["choices"][0]["message"]["content"].strip()

def generate_prompt(message):
  messages = [{"role": "system", "content": "You are a helpful assistant."}]
  messages.append({"role": "user", "content": message})
  return messages

form = cgi.FieldStorage()
prolog_message = urllib.parse.unquote_plus(form.getvalue("message"))
messages = generate_prompt(prolog_message)
result = send_openai_request(messages)

print("Status: 200 OK")
print("Content-Type: text/html")
print("")
print(result)
```

## Prolog Program (prolog_chatgpt.pl)
* WEB_SERVER_URL must be replaced by the URL of the server that hosts the script gpt-prolog.py.

```prolog
:- use_module(library(http/http_client)).
:- op(900,fy,not).

% CALLING ChatGPT

chatgpt_api_url('WEB_SERVER_URL/gpt-prolog.py').

ask_chatgpt(Msg,Reply) :-
   chatgpt_api_url(Url),
   http_post(Url,
     form_data([ message = Msg]),
     Reply,
     [cert_verify_hook(cert_accept_any)]).

% TELLING A STORY

narrate_story(St,Nstl) :-
  prefix_story(St,Stp),
  term_string(Stp, Stps),
  prompt_sto(P),
  string_concat(P, Stps, Msg),
  ask_chatgpt(Msg,Nst),
  text_list(Nst,Nstl).
```

```prolog
prompt_sto(P) :-
  P = 'Please narrate the plot:'.

narrate_events(St,Nstl) :-
  prefix_story(St,Stp),
  term_string(Stp, Stps),
  prompt_evs(P),
  string_concat(P, Stps, Msg),
  ask_chatgpt(Msg,Nst),
  text_list(Nst,Nstl).

prompt_evs(P) :-
  P = 'Please narrate separately each event of the following plot, skipping a line after the narrative of each event: '.

prefix_story(St, Stp) :-
  findall(Evp,
    (member(Ev,St),
     prefix_parameters(Ev, Evp)),
    Stp).

prefix_parameters(Ev, Evp) :-
  Ev =.. [Op|L],
  entities(El),
  findall(Pref: Parm,
    (member(Parm, L),
     member(Pref, El),
     Ep =.. [Pref,Parm],
     Ep),
    Lp),
  Evp =.. [Op|Lp].

text_list(T,Lt) :-
  name(T,L1),
  t(L1,L2),
  v(L2,Lt).

t([A],[]).
t([13,10,13,10|R],[13|S]) :- t(R,S).
t([13,10|R],[13|S]) :- t(R,S).
t([A|R],[A|S]) :- t(R,S).

v([],[]).
v(L,[T|S]) :- append(A,[13|B],L), name(T,A), v(B,S).

% EXAMPLE DOMAIN

entities([hero, villain, victim, sentinel, place, crime]).

% entity instances

 hero('Sir Brian').
 villain('Draco').
 victim('Princess Marian').
 sentinel('Walt').
 place('White Castle').
 place('Fortress of Draco').
 crime(capture).

% TESTS

test1(Lst) :- story(St), narrate_story(St,Lst).

test2(Lst) :- story(St), narrate_events(St,Lst).

 story(St) :- St =
```

```
[ride('Draco', 'White Castle'),
 overpower('Draco','Walt'),
 capture('Draco','Princess Marian'),
 carry('Draco','Princess Marian','Fortress of Draco'),
 warn('Walt','Sir Brian','capture'),
 ride('Sir Brian','Fortress of Draco'),
 fight('Sir Brian','Draco'),
 defeat('Sir Brian','Draco'),
 free('Sir Brian','Princess Marian'),
 carry('Sir Brian','Princess Marian','White Castle')].
```